# Nuclear Astrophysics with γ-ray Line Observations

**Roland Diehl[1]**
*Max Planck Institut für extraterrestrische Physik*
*D-85748 Garching, Germany*
*E-mail:* `rod@mpe.mpg.de`

Gamma-ray spectrometers with high spectral resolution have been operated in space since 2002. Major efforts to understand instrumental response and backgrounds are a requird before detailed science interpretations can be derived; by now, high-resolution line-shape studies have resulted in significant astrophysical constraints, not only through studies of solar-flare details, but also for nuclear processes in the Galaxy: $^{44}$Ti from the Cas A supernova could only be detected in the low-energy lines at 68 and 78 keV, the 1157 keV line from the same decay is not seen; this constrains $^{44}$Ti ejection in core collapse supernovae. Diffuse nucleosynthesis is studied through $^{26}$Al, $^{60}$Fe, and positron annihilation gamma-ray measurements. The gamma-ray line from decay of radioactive $^{26}$Al could be measured at unpredecented spectroscopic precision. The new determination of the total mass of $^{26}$Al produced by stellar sources throughout the Galaxy yields 2.8 ±0.9 $M_\odot$, and the interstellar medium around $^{26}$Al sources appears characterized by velocities in the ~100 km s$^{-1}$ region. $^{60}$Fe is clearly detected with SPI, its intensity ratio to $^{26}$Al of ~15% is on the lower side of predictions from massive-star and supernova nucleosynthesis models. Nucleosynthesis sources are probably minor contributors to Galactic positrons; this may be deduced from the bulge-centered spatial distribution of the annihilation gamma-ray emission, considering that nucleosynthesis sources are expected to populate mainly the disk part of the Galaxy. It is evident that new views at nuclear and astrophysical processes in and around cosmic sources are being provided through these space missions.

---

[1] Speaker

# 1. Gamma-ray Spectrometers in Space

Cosmic nuclei can be observed near their formation sites most directly when they emit characteristic gamma-ray lines from de-excitation, e.g. after radioactive decay. Alternative measurements of cosmic nuclei are more indirect; examples are stellar photospheric absorption lines, or mass spectrometry of meteoritic inclusions. Gamma-ray measurements and meteoritic laboratory analyses provide isotopic information, while only elemental information can be obtained in most spectroscopic methods. Nuclear-reaction processes and the physics of nuclear burning regions inside these sources are best constrained through isotopic abundance data. Yet, the technique of gamma-ray telescopes is complex, and less precise than other cosmic abundance measurements, mainly from two reasons: Spatial resolutions of ~degrees and signal-to-background ratios of ~% restrict contributions from gamma-ray astronomy to nearby sources in the Galaxy (up to 10 Mpc for SNIa $^{56}$Ni radioactivity). The advantages of gamma-ray astronomical data (*isotopic* information, unaffected by physical conditions in/around the source such as temperature or density, and nearly un-attenuated along the line-of-sight due to their penetrating nature) makes such efforts worthwhile.

Recent launches of observatories with high-resolution solid state detectors (RHESSI [1], INTEGRAL/SPI [2,3,4], Figure 1) into space have added a new quality to this field since 2002: Spectroscopy of these nuclear lines allows to identify cosmic lines against background references, and is capable to constrain the kinematics of cosmic isotopes in the gamma-ray

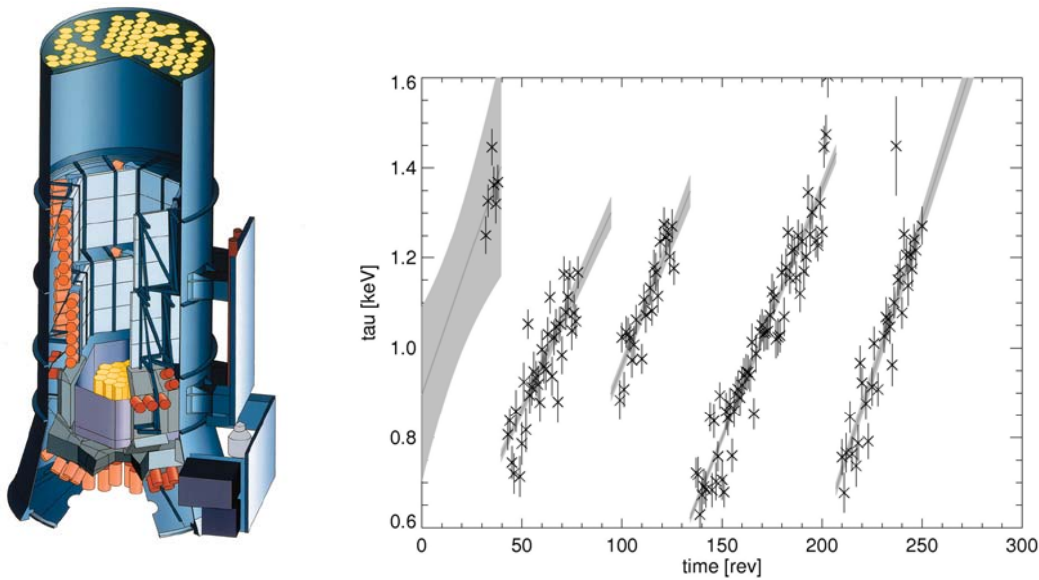

**Figure 1:** The SPI instrument[3,4] is built around a 19-element camera of Ge detectors; incident gamma-rays will cast a characteristic shadow onto this camera due to the coded mask, which allows to discriminate sources against instrumental background (left). Cosmic-ray bombardment in space destroys the charge-collection properties of the Ge detectors. Periodic annealing cures these defects, such that the high spectral resolution can be maintained over years. Shown is the degradation parameter τ versus time in units of INTEGRAL's 3-day orbits[2] (right)

emission region. The astrophysical importance of such velocity information for the study of particle acceleration in solar flares has been demonstrated and will not be further addressed here; but it will be demonstrated below for cosmic sources, i.e. for supernova nucleosynthesis ejecta and for the turbulent interstellar medium around massive stars. The special processes involved in positron-annihilation gamma-ray production[5] shape the corresponding gamma-ray line at 511 keV. This can be exploited to constrain ionization state and temperature of the interstellar gas in the region where positrons annihilate.

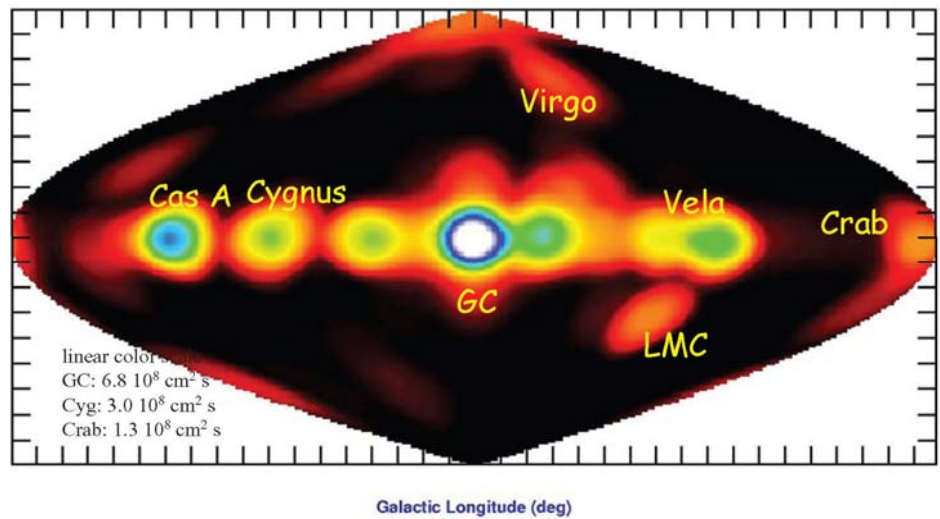

**Figure 2:** INTEGRAL/SPI exposure from more than 3 years of data. This all-sky map in Galactic coordinates sums up orbits 16-437 (Nov 30, 2002-May 15, 2006). Prominent source regions are labelled.

Candidate sources of characteristic gamma-ray lines are supernovae and novae, but also the winds from massive stars[6,7]: Freshly-produced nuclei from explosive layers near the surface of compact stars and from stellar interiors are ejected into interstellar space, where their decay can be observed directly in gamma-rays. Once ejected, long-lived radio-isotopes will decay in interstellar space; once merged with interstellar gas, they provide radioactive tracers which reflect the state of circumstellar and interstellar gas around the nucleosynthesis source. The kinematics of inner ejecta (rather than outer envelope and swept-up matter) can be studied in young supernova remnants, more-or-less complete ionization of matter will lead to modified decay histories for the total of $^{44}$Ti produced in the explosion (decay time 85 years, decay by electron capture only). Longer-lived isotopes such as $^{26}$Al ($\tau\sim1.04$ Myrs, [8]) will reflect interstellar-gas kinematics through Doppler shifts or broadenings of decay gamma-ray line(s); this interstellar medium around massive star sources is otherwise hard to study, as it is hot and tenuous in nature. Positrons from decays of isotopes on the proton-rich side of the valley of isotopic stability will typically be produced at MeV energies[5]; their propagation before their annihilation is more complex, being controlled by density and magnetic field morphologies near the nucleosynthesis sources[5,9]; high-energy emission from annihilation-in-flight would be the 'smoking gun' near the source, while the 511 keV line and lower-energy continuum arise from thermalized positrons with therefore-unknown origin.

In this paper, we will discuss what has been learned from RHESSI and INTEGRAL with respect to nucleosynthesis sources in our Galaxy, after these experiments have been studied in detail with respect to their responses and backgrounds and can be considered mature. Initial results had been reported, a.o., at NIC XIII [10]; the sky exposure of INTEGRAL is shown in Figure 2.

## 2. $^{44}$Ti from Cas A: Interpreting the Missing Signal

The Cas A supernova event has provided us with one of the best opportunities to study a young (~340 years old) and relatively-nearby (~3.4 kpc) supernova remnant clearly originating from gravitational collapse of a massive star. Ejecta kinematics has been studied through recombination lines in great detail, and the emission morphology of the remnant imaged in X-rays through optical to infrared emission suggests substantial mixing [11]. After the pioneering discovery of $^{44}$Ti radioactivity gamma-rays with COMPTEL [12], several other instruments now have confirmed the signature of $^{44}$Ti decay: First, Beppo-Sax [13] found emission from the 68 and 78 keV lines from $^{44}$Sc deexcitation, and now INTEGRAL/IBIS also clearly detected those $^{44}$Sc hard X-ray lines [14]. Taken together, the gamma-ray intensity measurements from several instruments suggest a gamma-ray flux of ~2.5 (±0.5) $10^{-5}$ ph cm$^{-2}$ s$^{-1}$ (Figure 3). Depending on distance, radioactive lifetime, and ionization state for $^{44}$Ti, this translates into a $^{44}$Ti yield of 0.8-2.5 $10^{-4}$ M of $^{44}$Ti produced in the supernova event. Comparing this to model predictions, Cas A's $^{44}$Ti yield appears somewhat on the high side. Note that a recent refinement of the $^{44}$Ti lifetime measurement specifies this to value of 85.2 ±0.3 years [15], from following the decay curve over 12 years.

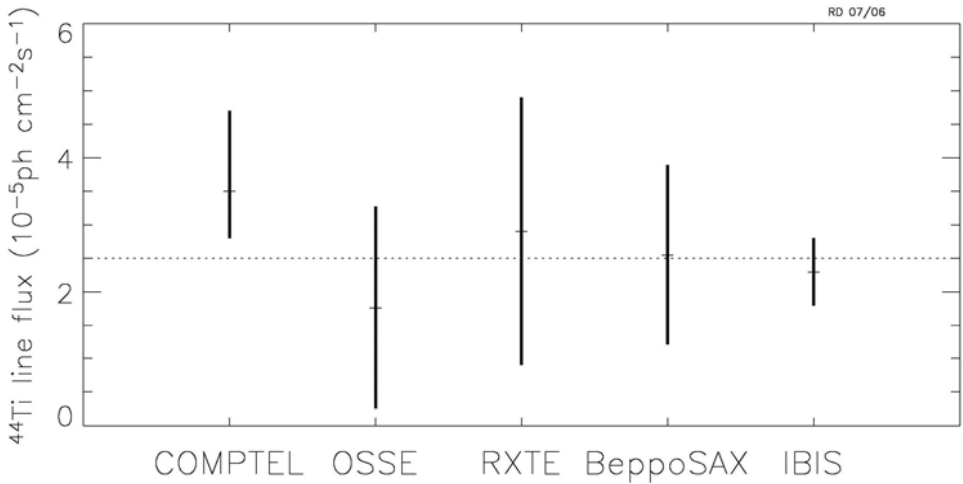

**Figure 3:** Gamma-ray flux from Cas A from radioactive decay of $^{44}$Ti, as measured by several X/gamma-ray instruments

Increased $^{44}$Ti production in Cas A due to deviations from spherical symmetry in the explosion had been suggested, combining the evidence from observations of Cas A's jet )see e.g. [16]) and fast-moving knots (see e.g. [17]) with supernova explosion models where polar regions show higher entropy caused by presupernova stellar rotation. This is reinforced by comparisons of $^{56}$Ni and $^{44}$Ti yields for SN1987A, and by conversion of the solar $^{56}$Fe to $^{44}$Ca abundance ratio [18,19]: In all cases, the $^{44}$Ti yields of 1D models for supernova-production relative to $^{56}$Ni appear lower than indicated by these measurements by a factor up to three. Note however that and a recent re-determination of the key production reaction $^{40}$Ca($\alpha,\gamma$)$^{44}$Ti suggests a factor ~2 increase in model yields [20], weakening such arguments.

No other core-collapse supernova has been seen clearly yet in $^{44}$Ti radioactivity, despite extensive searches with hard-X and gamma-ray telescopes (tentative signals had beenreported for a source in the Perseus region and for the Vela Junior remnant; none of these could be confirmed by later observations of the same instrument nor by independent observations). This is a remarkable result: It implies that something is odd in our expectations, which would yield typically a few observable sources from the massive-star population of the inner Galaxy [18]. The obvious conclusion is that $^{44}$Ti ejection is not common among core-collapse supernovae, and Cas A an exception rather than the rule. 3D effects of the supernova explosion are the plausible cause of such a paucity of $^{44}$Ti sources.

The non-detection of the 1157 keV gamma-ray line with SPI fits into this picture: If $^{44}$Ti ejection occurs jet-like rather than in the form of the slowest inner part of ejecta, velocities are higher on average than the ~1000 km s$^{-1}$ typical for inner ejecta; above typical velocities of 7000-8000 km s$^{-1}$, the $^{44}$Ti line at 1157 keV would dissappear in the instrumental background of SPI's Ge detector signals [21]. Improved search for this high-energy line with SPI are being made, to consolidate this important constraint.

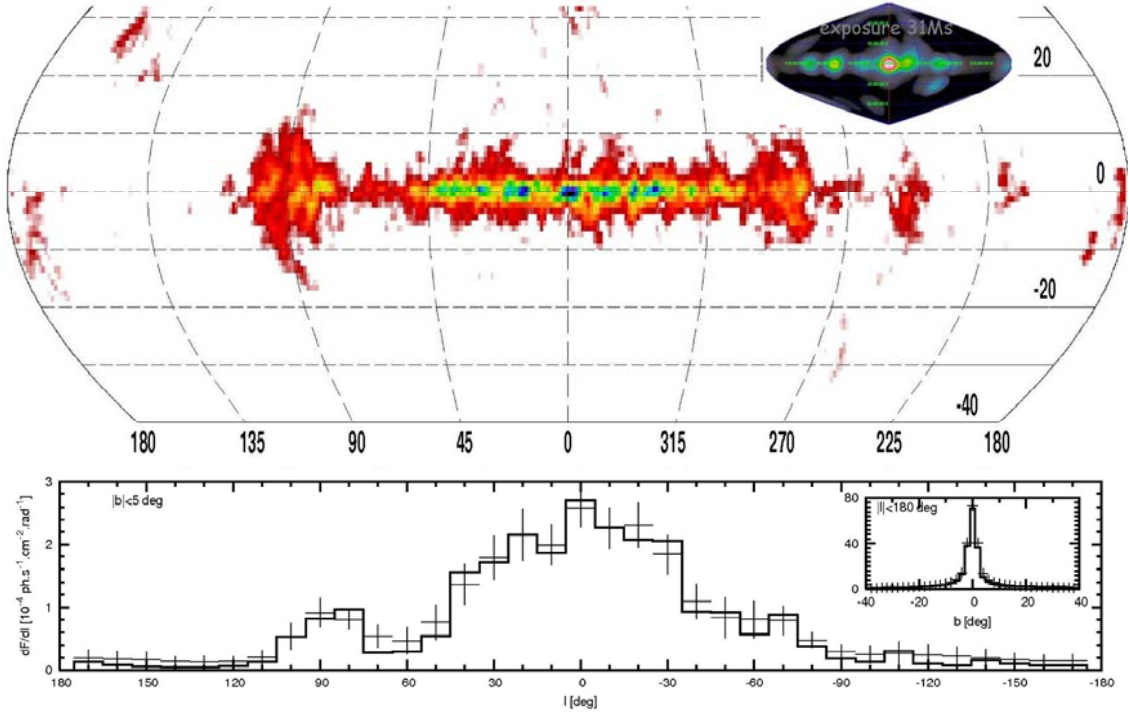

**Figure 4:** Imaging deconvolution of INTEGRAL/SPI data from a 6 keV-wide energy band including the 1808.65 keV line from $^{26}$Al. The inset shows the exposure of observations used for this all-sky map, presented in Galactic coordinates. Profiles in Galactic longitude and latitude (inset) are shown below the map. (Halloin et al., in preparation for A&A)

## 3. Sources of $^{26}$Al in the Galaxy: New Diagnostics of Galactic Kinematics

The 1808.65 keV gamma-ray emission from the decay of $^{26}$Al in the Galaxy [8] has been mapped by COMPTEL from 9 years of observations [22]; its disk-like distribution with irregularities and prominent intensity in the Cygnus region argues for predominant production through massive stars. Two years of data from INTEGRAL/SPI have now resulted in first images in this gamma-ray line (Figure 4), using SPI's coded-mask imaging at a limiting spatial resolution of ~2.7° (cmp. COMPTEL: ~3.8°). Although still noisy, the basic characteristic emission features are confirmed: $^{26}$Al emission extends along the Galactic plane, with asymmetries (the fourth Galactic quadrant being brighter than the first) and with prominent emission from the Cygnus region.

Already from first spectroscopy of the $^{26}$Al line from the inner Galaxy with space-borne Ge detector instruments, the line was found to be rather narrow [23,24], and not as broad as had been reported by an earlier Balloone-borne experiment [25]. With present data from SPI, any astrophysical broadening is found to be on the order of ~1 keV, with a firm upper limit (2σ) of 2.8 keV (Figure 5, Figure 6; from [26]). Therefore, line broadenings as expected from large-scale Galactic rotation and from interstellar turbulence in massive-star regions are fully compatible with these recent measurements.

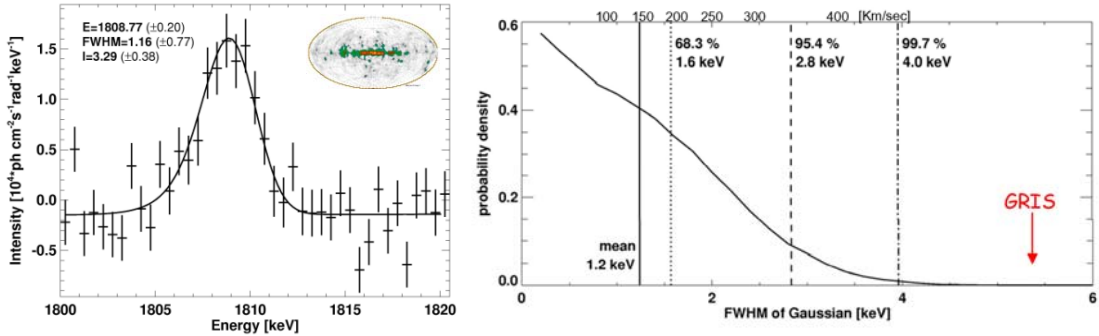

**Figure 5 (left):** The SPI measurement of the $^{26}$Al line from the inner Galaxy finds the line broadening from astrophysical effects (Galactic rotation, ISM turbulence) to be rather small [26].

**Figure 6 (right):** SPI constraints on the line broadening: A previously-reported broad line (corresponding to ISM velocities of 540 km s$^{-1}$ [25], labelled 'GRIS') can be excluded, lying in the far tail of the probability distribution; velocities of ~100 km s$^{-1}$ are plausibly consistent with the data.

With fine spectroscopy and deep exposure of the inner region of the Galaxy, systematic centroid shifts of the $^{26}$Al gamma-ray line were seen, consistent with what is in principle expected from the differential rotation of the Galactic disk. Line centroid offsets are few tenths of keV only, hence require the maintenance of fine energy resolution over these multi-year exposures through annealing, and careful accounting of the time-variable spectral resolution due to cosmic ray damages (cmp. Figure 1). From this signature of Galactic rotation, one may conclude that the observed source regions are located in the inner regions of the Galaxy indeed, and that therefore the observed $^{26}$Al gamma-ray flux represents $^{26}$Al emission from all sources integrated over the entire Galaxy. With this assumption, and a geometrical model for the Galaxy's massive-star population along the Galactic plane, the observed $^{26}$Al line flux translates into a total Galactic amount of 2.8 ±0.9 M$_\odot$, where the uncertainty mainly arises from different spatial distribution models for the $^{26}$Al sources [27]. Because the $^{26}$Al origin is predominantly attributed to massive stars, this $^{26}$Al amound can be converted into a core-collapse supernova rate, assuming a steady-state situation and using the $^{26}$Al production per stellar mass from nucleosynthesis models with the standard mass distribution function for massive stars (for details see [27]). Therefore, a new and rather direct measurement of this key parameter of our Galaxy, the rate of stellar births and deaths, can be employed with penetrating gamma-rays, from the otherwise occulted massive-star population in the inner regions of our Galaxy; the INTEGRAL/SPI-derived value is 1.9 (±1.1) supernovae yr$^{-1}$, corresponding to a conversion of ~4 M$_\odot$ of interstellar gas into stars per year, galaxy-wide.

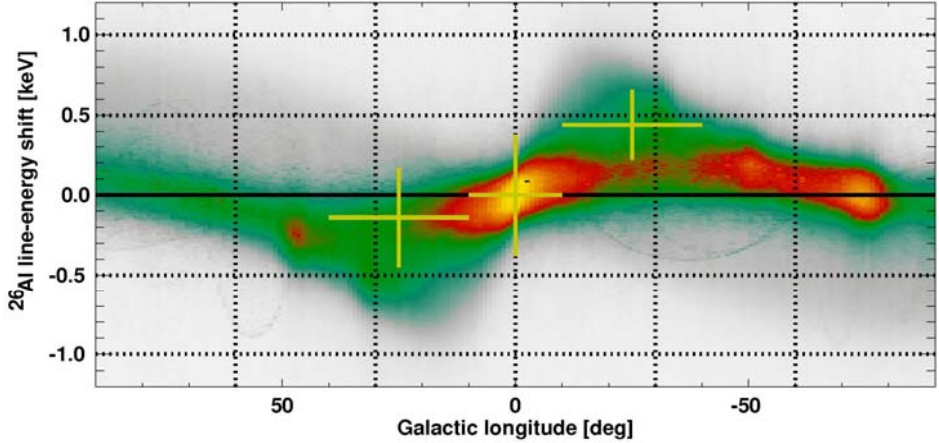

**Figure 7:** Centroid shifts of the $^{26}$Al line along the plane of the Galaxy (data points) are consistent with the pattern expected from Galactic rotation (color scale) [27].

## 4. Annihilation of Positrons: Nucleosynthesis Origin or not?

Positrons have been seen to annihilate in our Galaxy, through their characteristic annihilation-line gamma-rays at 511 keV energy (for a recent review, see [7]). With INTEGRAL/SPI, for the first time this gamma-ray emission could be mapped over large regions of the sky [28,29], which resulted in a surprise that puzzles a broad community of astrophysicists (see detailed discussion and references in [29]): Clearly, nuclear processes in stars, supernovae, and novae are expected to release positrons from radioactive decays of isotopes on the proton-rich side of the valley of stable isotopes. Also, electron-positron pair-production is likely near compact stars, either because strong magnetic fields of rapidly-rotating neutron stars result in particle-accelerating potential gaps, or accretion flow in a binary system onto its compact-star component results in plasma jets such as observed in microquasars. All such stellar sources are expected to release positrons into interstellar space in their vicinity, and thus positrons should be abundant in the stellar disk and bulge of the Galaxy. The surprise is that annihilation emission appears to arise from the central region only, or at least much brighter and outshining a Galactic-disk component. It seems hard to imagine that some transport process could collect all positrons in the Galaxy's central region, although the positron lifetime before annihilation can be as large as 10 My [5], and large-scale symmetry of the magnetic-field structure in the Galaxy's halo may be provide a path for long-lived positrons into the Galaxy's bulge region (e.g. [30]). In the absence of a plausible explanation, a variety of speculative suggestions have been made, such as major positron production from annihilation of light dark-matter particles, from accretion flows onto the Galaxy's central supermassive black hole, or from very energetic supernovae and gamma-ray bursts (see [29] for discussion and references. Although the total intensity of the annihilation gamma-rays is consistent with the positron budget that may be attributed to nucleosynthesis and other stellar production, the sky image of annihilation emission appears to clearly exclude a disk-like young stellar population as main positron source, provided annihilation occurs not too far from the sources. Determining the positron annihilation rate near such presumed sources appears to be the prominent experimental challenge for a next generation of gamma-ray telescopes. INTEGRAL/SPI spectroscopy of the 511 keV line and the 3-photon annihilation continuum from annihilation through formation of intermediate positronium atoms has already constrained the annihilation medium to be partially-ionized and at intermediate ISM temperatures around 8000 K [31,32]; the propagation of positrons from their sources into their annihilation sites will be an interesting astrophysical problem, to be addressed by such gamma-ray measurements, and helping to clarify the origina and propagation of cosmic rays in the Galaxy.

## 5. $^{60}$Fe Production by Supernovae: A Consistent Picture?

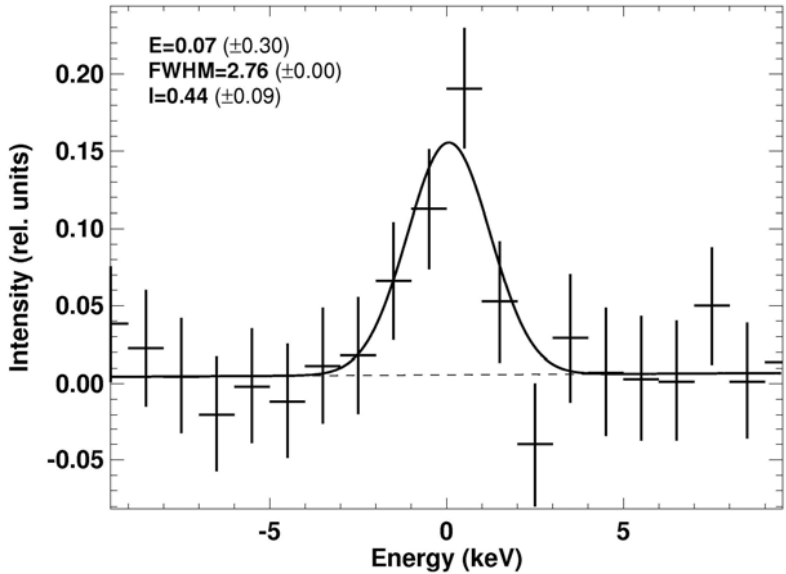

**Figure 8:** INTEGRAL/SPI $^{60}$Fe signal from the inner Galaxy [41]. Data from both lines have been superimposed, referencing to the laboratory line energies of 1173.23 keV and 1332.49 keV, respectively. The solid line represents a Gaussian fit with its width fixed from nearby instrumental background lines.

Stellar nucleosynthesis models have always predicted that the convective envelopes of massive-stars are plausible production sites of long-lived $^{60}$Fe isotopes ($\tau\sim$2.2 My), which would be ejected in the terminal core-collapse supernova explosion [38, 33, 34,35]. Because of the dominant massive-star origin of $^{26}$Al, the gamma-ray brightness ratio $R_{Fe-Al}=I_\gamma(^{60}Fe) / I_\gamma(^{26}Al)$ from the emission of diffuse nucleosynthesis in steady-state over the last several My is a suitable test of these massive-star nucleosynthesis models.

With RHESSI, for the first time a clear hint for $^{60}$Fe gamma-rays had been recognized in 2003 [36]. But the signal is weak, the gamma-ray emission was found to be 16 (±5)% of $^{26}$Al brightness [37]. This seemed to confirm earlier predictions of ~15% [38] very well, but it was clear several later studies of massive-star nucleosynthesis that substantial uncertainties both from stellar structure (establishment of suitable convective-burning regions) and from nuclear cross sections (n capture on unstable $^{59}$Fe is difficult to measure, its β decay lifetime is uncertain) could also result in substantially higher model-predicted $^{60}$Fe/$^{26}$Al brightness ratios $R_{Fe-Al}$ (see discussion of studies since 1995 by [39]).

Confirmation of the RHESSI $^{60}$Fe signal was reported from first-year INTEGRAL/SPI data, although the overall significance ws marginal (~3σ) and systematics due to instrumental background lines was a concern [40]. Now, a recent re-analysis with more data finds a significant $^{60}$Fe signal [41] (at 5σ), instrumental systematics could be reduced, and the $^{60}$Fe signal appears consistently in both the $^{60}$Fe decay gamma-ray lines at 1173 and 1332 keV and in both single-detector and multiple-detector hits of the SPI Ge camera. The new SPI data find a $^{60}$Fe/$^{26}$Al gamma-ray brightness ratio $R_{Fe-Al}$ of 14%, with an uncertainty of a few percent (detailed assessment is in progress, [41]).

## 6. Summary and Prospects

The new high-resolution spectroscopy of cosmic gamma-rays with space-borne Ge detectors has demonstrated the power of gamma-ray lines from nuclear and high-energy nonthermal processes for astrophysics: $^{44}$Ti, $^{26}$Al and $^{60}$Fe radioactive decay has been measured from cosmic sources and reflects nuclear reactions inside massive stars. With such spectroscopy, width and centroid constraints for $^{44}$Ti constrains the kinematics of inner ejecta from core-collapse supernovae; similarly, and the line shape and centroid of the $^{26}$Al line reflects nucleosynthesis ejecta and their dynamics in otherwise hardly-accessible hot and tenuous gas regions near nucleosynthesis sources. With $^{60}$Fe gamma-rays and their ratio to $^{26}$Al, the models for massive-star nucleosynthesis can be tested in a global sense. Although it is difficult to combine imaging and spectroscopy in an astronomical instrument for gamma-rays, the INTEGRAL/SPI results demonstrate that such combination may reveal new frontiers, as in the case of the positron annihilation gamma-ray line with its puzzling morphology dominated by an extended inner-Galaxy source. INTEGRAL will continue to operate well into the next decade, and thus can provide a gamma-ray line survey of the Galaxy down to the level of $10^{-5}$ ph cm$^{-2}$ s$^{-1}$. This can address only the brightest sources of nuclear lines, but this prepares the ground for deeper exploration of cosmic sources in emission which originates in atomic nuclei.

**Acknowledgements.** This contribution results from the collaborative work with the members of the INTEGRAL and SPI Teams, and fruitful discussions with many other colleagues; I am grateful for their collaboration, in particular to Hubert Halloin, Karsten Kretschmar and Andrew Strong at MPE, Pierre Jean, Jürgen Knödlseder, and Jean-Pierre Roques at CESR, Trixi Wunderer at SSL Berkeley, Bonnard Teegarden at GSFC Greenbelt, Jacco Vink at SRON Utrecht, Nikos Prantzos at IAP, Marco Limongi and Alessandro Chieffi at CNR Frascati, Dieter Hartmann at Clemson University, and Stan Woosley at UC Santa Cruz. INTEGRAL is an ESA project with instruments and science data centre funded by ESA member states (especially the PI countries: Denmark, France, Germany, Italy, Switzerland, Spain), Czech Republic and Poland, and with the participation of Russia and the USA. The SPI spectrometer has been completed under the responsibility and leadership of CNES/France, its anticoincidence system is supported by the German government through DLR grant 50.0G.9503.0. We acknowledge the support of INTEGRAL from ASI, CEA, CNES, DLR, ESA, INTA, NASA and OSTC.